\documentclass[aps,prd,preprint,superscriptaddress,tightenlines,nofootinbib]{revtex4}

\usepackage{graphicx}
\usepackage{dcolumn}
\usepackage{bm}

\begin{document}

\preprint{CLEO CONF 07-1}   

\title{Dalitz plot analysis of the $D^+\to K^-\pi^+\pi^+$ decay}
\thanks{Submitted to The 2007 Europhysics Conference on High Energy Physics,
        July 19-25, 2007, Manchester, England.}


\author{G.~Bonvicini}
\author{D.~Cinabro}
\author{M.~Dubrovin}
\author{A.~Lincoln}
\affiliation{Wayne State University, Detroit, Michigan 48202, USA}
\author{D.~M.~Asner}
\author{K.~W.~Edwards}
\author{P.~Naik}
\affiliation{Carleton University, Ottawa, Ontario, Canada K1S 5B6}
\author{R.~A.~Briere}
\author{T.~Ferguson}
\author{G.~Tatishvili}
\author{H.~Vogel}
\author{M.~E.~Watkins}
\affiliation{Carnegie Mellon University, Pittsburgh, Pennsylvania 15213, USA}
% ...
\collaboration{CLEO Collaboration} 
\noaffiliation


\date{July 19, 2007}

\begin{abstract} 
We present a Dalitz plot analysis of the decay $D^+ \to K^- \pi^+ \pi^+$
based on 281~pb$^{-1}$ of $e^+e^-$ collision data produced at the $\psi(3770)$ by CESR and
observed with the CLEO-c detector. 
We select 67086 candidate events with a small, $\sim$1.1\%, background
for this analysis.
When using a simple isobar model our results are consistent with the
previous measurements done by E791. 
Since our sample is considerably larger we can explore
alternative models.  We find better agreement with data when we
include an isospin-two $\pi^+\pi^+$ S-wave contribution.
We apply a quasi model-independent partial wave analysis and
measure the amplitude and phase of the $K\pi$ and  $\pi^+\pi^+$ S waves in the range
of invariant masses from the threshold to the maximum in this decay.
\end{abstract}

\maketitle

\newcommand{\DECAY}{$D^+\to K^-\pi^+\pi^+$}
\newcommand{\PM}{$\pm$}
\newcommand{\FIGDIR}{.}
\newcommand{\FIGDIRSEL}{.}
\newcommand{\FIGDIRW}{.}

In comparison to many other $D^+$ meson modes, the $D^+ \to K^-\pi^+\pi^+$ decay is unique
in many aspects.
A large contribution of over 60\% from a $K\pi$ S-wave intermediate state has been 
observed in earlier 
experiments~\cite{MARK-III-1987,NA14-1991,E691-1993,E687-1994,E791-2002,E791_Kpipi}.
The large branching fraction, ${\cal B}(D^+ \to K^- \pi^+ \pi^+)=(9.51 \pm 0.34)\%$, for
this mode makes it the usual choice for normalization of other $D^+$ meson decay rates.
Understanding its peculiar intermediate substructure will be very beneficial.
Two identical pions in this decay should obey Bose symmetry.
Assuming quasi two-body nature of this decay, it is dominated by the 
two identical sets of $K^-\pi^+$ waves interfering with each other. 
This symmetry significantly reduces the degrees of freedom in the regular Dalitz plot analysis 
and allows us to apply a model-independent partial wave analysis.
One would also expect a small contribution from the isospin-two $\pi^+\pi^+$ S wave,
which can also be explored in this decay.

The $D^+ \to K^-\pi^+\pi^+$ decay has been previously studied 
with the Dalitz plot technique in many experiments, including 
MARK III \cite{MARK-III-1987},
NA14     \cite{NA14-1991},    
E691     \cite{E691-1993},    
E687     \cite{E687-1994}, and   
E791     \cite{E791-2002,E791_Kpipi}.   
To achieve good agreement with their data, E791 \cite{E791-2002} has included
a low-mass $K^-\pi^+$ resonance, $\kappa$, that significantly re-distributed 
all fit fractions observed in earlier experiments. Due to the difference in model,
the results of this high-statistics experiment have been excluded from 
the world average by the PDG.
Compared to previous experiments CLEO-c~\cite{CLEO-c} has many advantages.
Our sample is four times larger than E791,
and it is very clean; the background is about 1\% of the selected sample.
The invariant mass resolution in this three-track $D$-meson decay is very good, 
$\sim$2.5~MeV/c$^2$.
Our experimental conditions are similar to that of
MARK III, where $D$ mesons are produced with small momentum.
The other experiments, all fixed-target,
study the $D$ mesons produced with large momentum, typically tens of GeV/$c$.

Interest in this decay was outlined in
Refs.~\cite{Bugg-2005,Bugg-2006,Oller-2005},
which focus on a low-mass $K\pi$ S-wave contribution.
These authors reanalyze the E791 data with their own models.  
E791 reinterpreted their own data
with a model independent partial wave analysis \cite{E791_Kpipi},
and it is this approach we apply in our analysis with minor modifications.

\begin{figure}[!htb]
  \includegraphics[width=54mm]{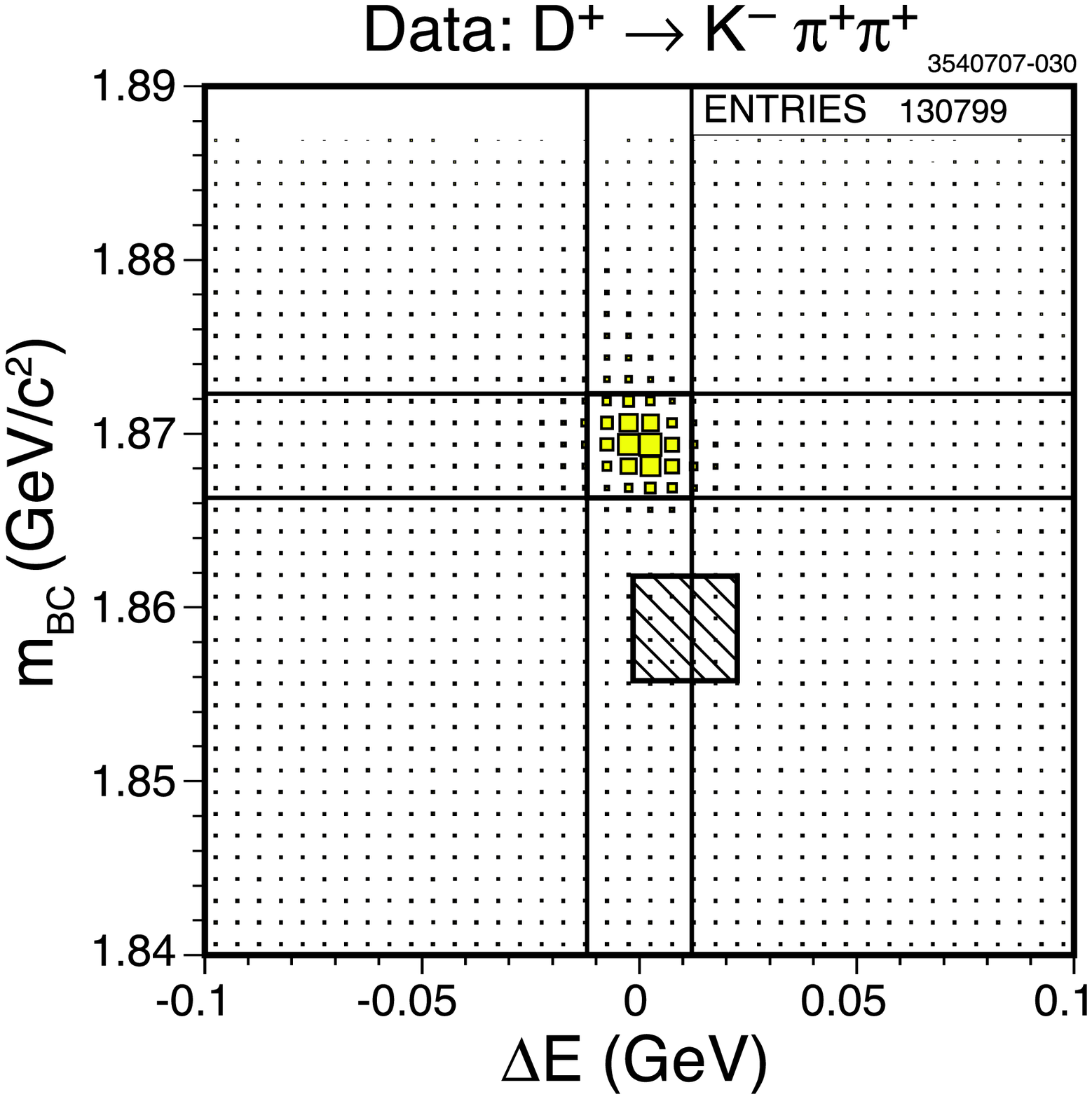} \hfill
  \includegraphics[width=54mm]{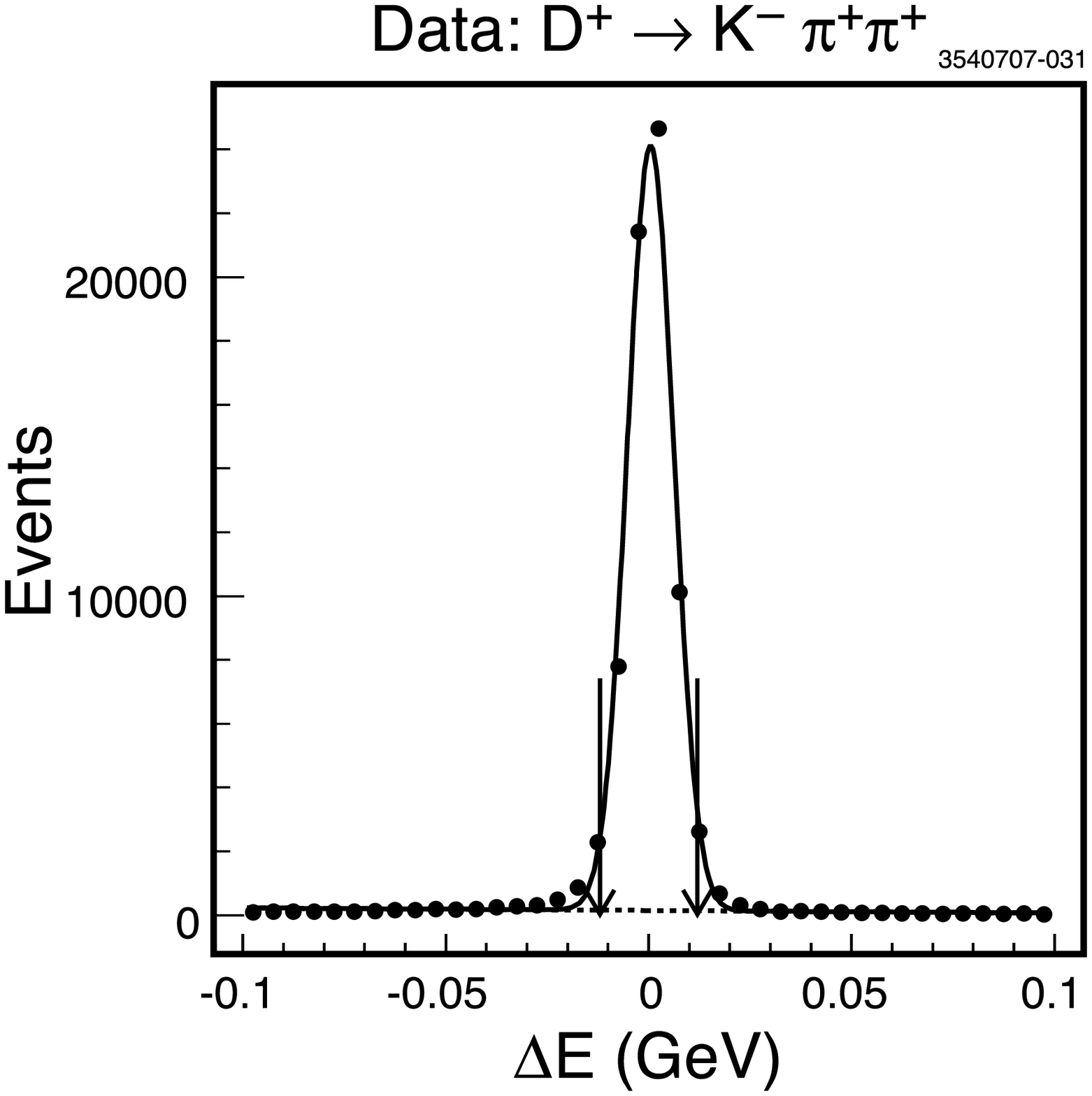} \hfill
  \includegraphics[width=54mm]{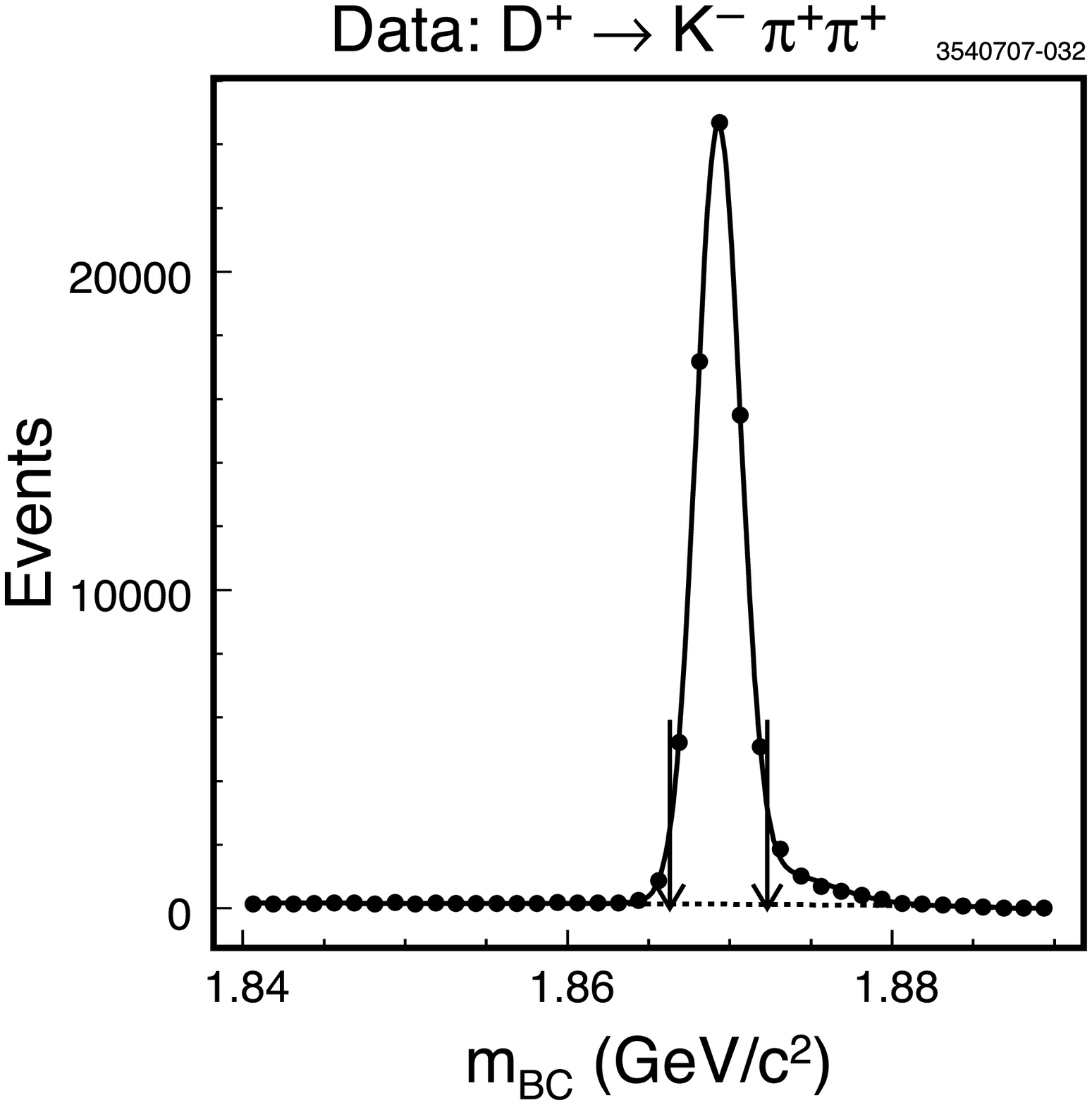}
  \caption{\label{fig:selection_1} Distribution of $m_{BC}$ {\it vs.} $\Delta E$ (left)
                  for $D^+ \to K^- \pi^+ \pi^+$ signal candidates in CLEO data. 
		  Events from the shown $\pm2\sigma$ bands crossing the signal region are
		  plotted as $\Delta E$ (middle) and $m_{BC}$ (right) distributions. 
             The background events are selected from the 
             sideband hatched box, which has about the same range of $K^- \pi^+ \pi^+$
             invariant mass as the signal box. 
          }
\end{figure}

We use 281~pb$^{-1}$ of $e^+e^-$ collisions at $\sqrt{s}\simeq 3770$~MeV,
produced by the Cornell Electron Storage Ring (CESR) and accumulated 
by the CLEO-c detector. This sample corresponds to
the production of $\sim$0.8~M $D^+D^-$ pairs in the process
$e^+e^- \to \psi(3770) \to D^+D^-$.
We reconstruct a one side $D$-meson decay using three tracks and applying 
the standard CLEO-c algorithms similar to Ref.~\cite{HadBF}.
In order to select $D^+ \to K^-\pi^+\pi^+$ decays,
we use two signal variables,
$\Delta E = E_D - E_{beam}$, and
$m_{BC} = \sqrt{E^2_{beam}-P^2_D}$,
where $E_{beam}$ is the beam energy, and
$E_D$ and $P_D$ are the energy and momentum of
the reconstructed $D^+$-meson candidate.
We require
$|\Delta E| < 2 \sigma(\Delta E)$ and
$|m_{BC} - m_D| < 2 \sigma(m_{BC})$, where resolutions
$\sigma(\Delta E) = 6$~MeV [(5.74$\pm$0.02)~MeV in single Gaussian fit], and
$\sigma(m_{BC}) = 1.5$~MeV/$c^2$ [(1.36$\pm$0.01)~MeV/$c^2$ in single Gaussian fit] 
represent the widths of
the signal peak in the 2D distribution and projections shown in
Fig.~\ref{fig:selection_1}.
We select 67086 candidate events for the Dalitz plot analysis \cite{Dalitz}.
The fraction of background, $\sim$1.1\%, in this sample is estimated from
the fit to $m_{BC}$ spectrum of Fig.~\ref{fig:selection_1}.

\begin{figure}[!htb]
  \includegraphics[width=52mm]{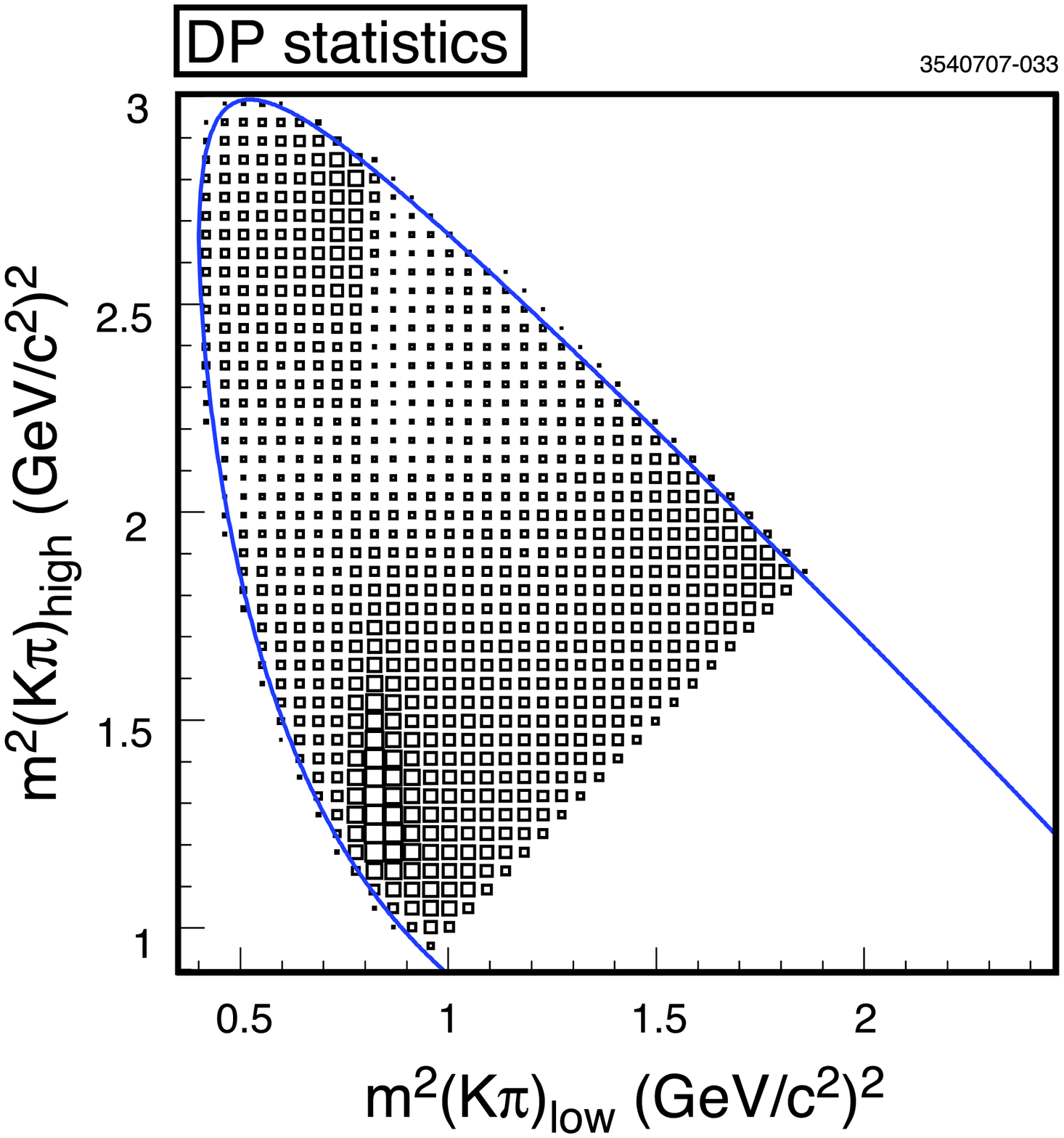} \hfill
  \includegraphics[width=54mm]{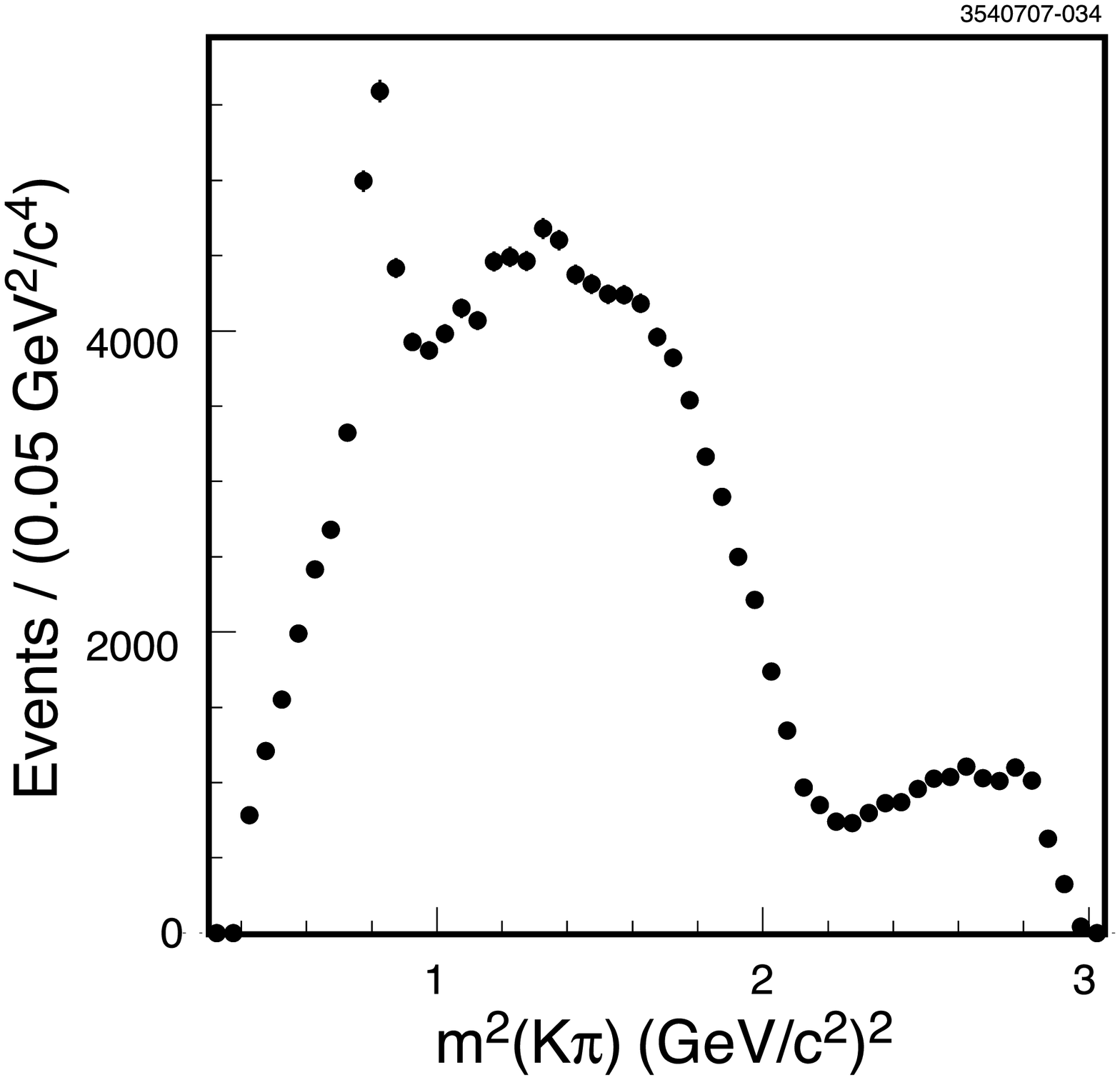} \hfill
  \includegraphics[width=54mm]{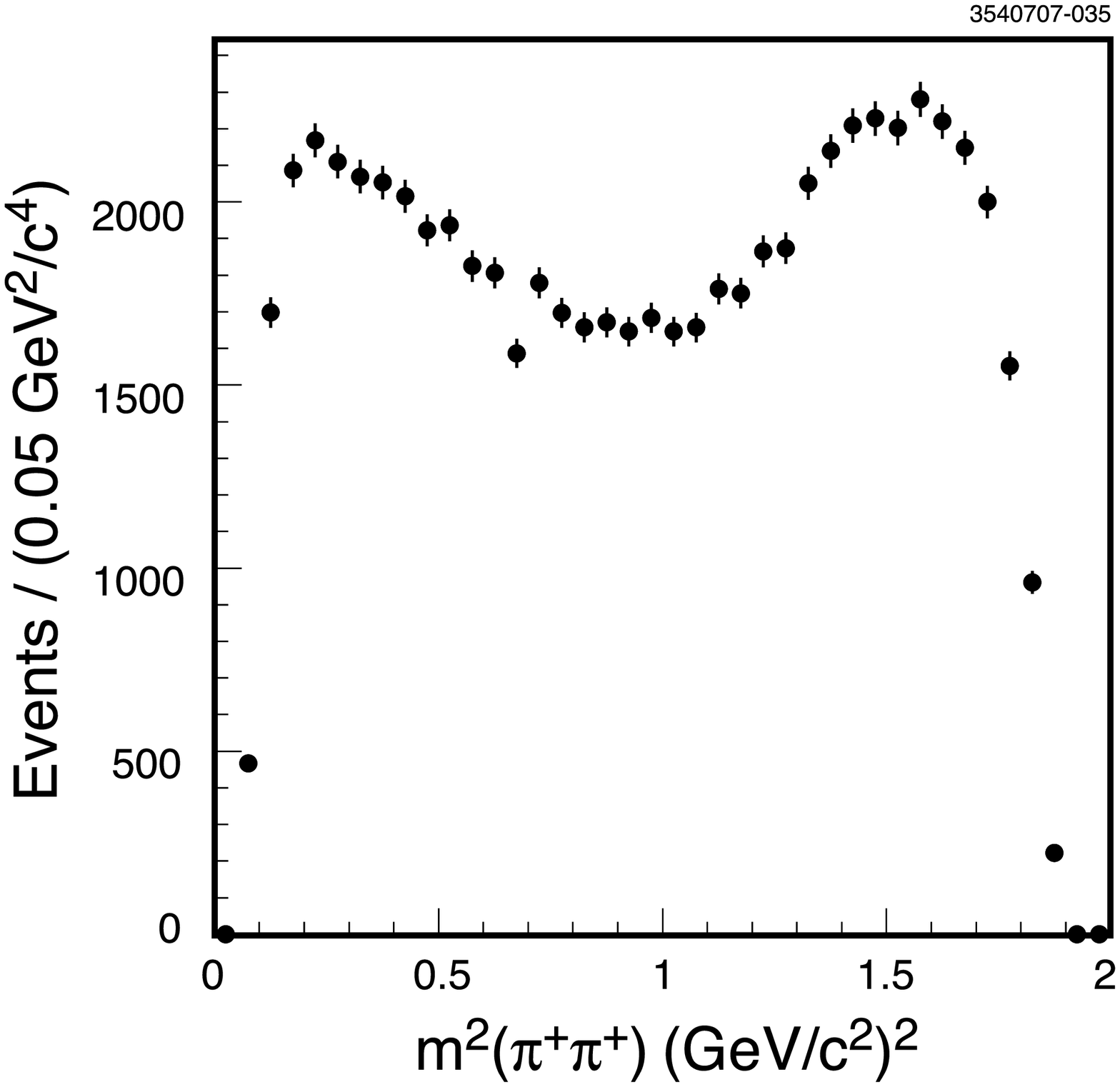}
  \caption{\label{fig:DP_data} Dalitz plot for data and its projections on 
            $m^2(K^-\pi^+)$(two entries per event) and $m^2(\pi^+\pi^+)$ axes.}
\end{figure}

In this analysis we apply a formalism similar to that 
used in other CLEO Dalitz plot analyses \cite{Tim,Dp3pi}.
The Dalitz plot is symmetric under the interchange of like-sign pions, 
so we do the analysis in the two dimensions of high $K\pi$
mass squared, $m^2(K\pi)_{high}$, versus
low $K\pi$ mass squared, $m^2(K\pi)_{low}$. 
This choice folds all data onto the top part
of the Dalitz plot, as shown in Fig.~\ref{fig:DP_data}(left).
Figure~\ref{fig:DP_data} also shows the $m^2(K\pi)$ (two entries per event) 
and $m^2(\pi\pi)$ projections.

The efficiency across the Dalitz plot is modeled with simulated events that are
fit to a two-dimensional third-order polynomial.  There is a notable
decrease of the efficiency in the corners of the Dalitz plot, which is modeled 
by the multiplicative sine-wave threshold functions.
The background shape is estimated from the $m_{BC}$ and $\Delta E$ sidebands.
Many possible resonances can contribute to the decay,
and a total of 7 different contributions are considered.  
Parameters describing these resonances are taken from previous experiments~\cite{PDG-2006}
or phenomenological publications.
Only contributions with an
amplitude significant at more than three standard deviations are said to
be observed.  Contributions that are not significant are
not included in the decay model used for the result.
We estimate a fit quality using Pearson's $\chi^2$
calculated for adaptive bins on the Dalitz plot.

First we compare our results, obtained in the framework of isobar model (IM),
with E791 Models A, B, and C from Ref.~\cite{E791-2002}. 
In particular, the most advanced Model C contains
$\bar K^*(892)\pi^+$,
$\bar K_0^*(1430)\pi^+$,
$\bar K^*(1680)\pi^+$,
$\bar K_2^*(1430)\pi^+$,  
$\bar K^*(1680)\pi^+$,
$\kappa\pi^+$, and non-resonant
contributions. 
Following E791 we allow an additional scalar $K\pi$ amplitude,
the ``$\kappa$'', as a Breit-Wigner with mass dependent width.
Gaussian form factors are used for $K_0^*(1430)$ and $\kappa$.
For all $K\pi$ resonances with non-zero spin,
the radii in the Blatt-Weisskopf~\cite{Blatt-Weisskopf}
form factors are fixed,
$r_D=5$~GeV$^{-1}$, $r_R=1.5$~GeV$^{-1}$.
Fit fractions obtained in our fit are compared with E791 in Table~\ref{tab:compare_results}.
We get a poor fit quality $\chi^2/\nu$=448/388, P($\chi^2,\nu$)=2\%;
the NR fit fraction is $\sim$10\%; and
the $\kappa\pi$ fit fraction is dominant, $\sim$30\%.
For the $K_0^*(1430)$ resonance we measure
$m_{K_0^*(1430)}=(1461\pm3)$~MeV/$c^2$ and
$\Gamma_{K_0^*(1430)}=(169\pm5)$~MeV/$c^2$,
which are consistent with E791 results,
but inconsistent with current PDG~\cite{PDG-2006} values.
Similar behavior is reported in a recent preprint 
from the FOCUS collaboration~\cite{FOCUS-2007}.
We also measure the $\kappa$ resonance parameters,
$m_{\kappa}=(805\pm11)$~MeV/$c^2$, and
$\Gamma_{\kappa}=(453\pm21)$~MeV/$c^2$.
\begin{table}[!htb]
\caption{\label{tab:compare_results} Comparison of fit fractions in
percent between this work and E791 results for an isobar and a QMIPWA model.
This work is preliminary and only statistical errors are shown.}
\begin{center}
\begin{tabular}{|l|c|c||c|c|}
\hline
     & \multicolumn{2}{|c||}{Model C} & \multicolumn{2}{|c|}{QMIPWA}               \\
\cline{2-5}
Mode &E791 \cite{E791-2002}& CLEO-c       &E791 \cite{E791_Kpipi} & CLEO-c         \\
\hline
NR   
     &  13.0\PM5.8\PM4.4   & 10.4\PM1.3   &  see S wave          & see S wave     \\ 
${\overline K}^*(892)\pi^+$		  
     &  12.3\PM1.0\PM0.9   & 11.2\PM1.4   &  11.9\PM0.2\PM2.0    & 10.0\PM0.3     \\ 
${\overline K}_0^*(1430)\pi^+$		  
     &  12.5\PM1.4\PM0.5   & 10.5\PM1.3   &  see S wave          & 11.4\PM3.6     \\ 
${\overline K}_2^*(1430)\pi^+$		  
     &  0.5\PM0.1\PM0.2    & 0.40\PM0.04  &  0.2\PM0.1\PM0.1     & 0.476\PM0.014  \\ 
${\overline K}^*(1680)\pi^+$		  
     &  2.5\PM0.7\PM0.3    & 1.36\PM0.16  &  1.2\PM0.6\PM1.2     & 2.52\PM0.08    \\ 
$\kappa \pi^+$				  
     &  47.8\PM12.1\PM5.3  & 31.2\PM3.6   &  see S wave          & see S wave     \\ 
\hline
Total S wave
     &  73\PM15            & 52\PM4       &  78.6\PM1.4\PM1.8    & 67.4\PM1.3     \\ 
\hline
$\chi^2 / \nu$, Prob.(\%) 
     & 46/63, 94\%         & 448/388, 2\% & 277/277, 47.8\%      & 368/346, 19.5\%\\ 
\hline
\end{tabular}
\end{center}
\end{table}
In these three models our fits
well reproduce behavior reported by the E791~\cite{E791-2002} experiment.
We get results for phases, fit fractions, and resonance parameters
which are statistically consistent with E791.
However the poor 2\%-probability of the fit motivates us
to explore an alternative model of the decay amplitude.

In an approach motivated by Ref.~\cite{Bugg-2006} we set the form factor
for any S-wave contribution to unity. 
Following prescriptions from Ref.~\cite{Oller-2005} we replaced the 
Breit-Wigner function for $\kappa$ by the complex pole approximation
${\cal  W}_{\kappa}(m)=[m^2_{\kappa} - m^2]^{-1}$, which is equivalent to the 
Breit-Wigner with a fixed width.  We find pole parameters,
$\Re m_\kappa=(651\pm17)$~MeV/$c^2$ and
$\Im m_\kappa=(-229\pm22)$~MeV/$c^2$, which are consistent with
the values obtained in Refs.~\cite{Bugg-2005} and \cite{Oller-2005}.  
We find no evidence of production or fit improvement by adding a 
$K^*(1410)\pi^+$ intermediate state.
We also tested the hypothesis~\cite{Bugg-2006} that the $K_0^*(1430)$ 
shape in this decay might be distorted by the vicinity of 
the $K\eta'$ threshold. We replace the Breit-Wigner function for $K_0^*(1430)$ 
by the Flatt\'e function~\cite{Flatte}:
\[{\cal  W}_{K_0^*(1430)} (m) = 
[m^2_{K^*(1430)} - m^2 - i( g^2_{K^*K\pi}   \rho_{K\pi}(m)
                     +g^2_{K^*K\eta}  \rho_{K\eta}(m)
                     +g^2_{K^*K\eta'} \rho_{K\eta'}(m)
                    )]^{-1},
\]
where $g^2$ and $\rho(m)=2P/m$ are the coupling constants and
the phase space factors, respectively.
We get the values of Flatt\'e parameters,
$m_{K^*(1430)}=(1468.3\pm4.2)$~MeV/$c^2$,
$g_{K^*K\pi}=(543\pm11)$~MeV/$c^2$,
$g_{K^*K\eta'}=(268\pm86)$~MeV/$c^2$, and
$g_{K^*K\eta}$ statistically consistent with zero,
and the fit quality is not meaningfully improved.


\begin{figure}[!htb]
 \begin{minipage}[t]{54mm}
  \includegraphics[width=53mm]{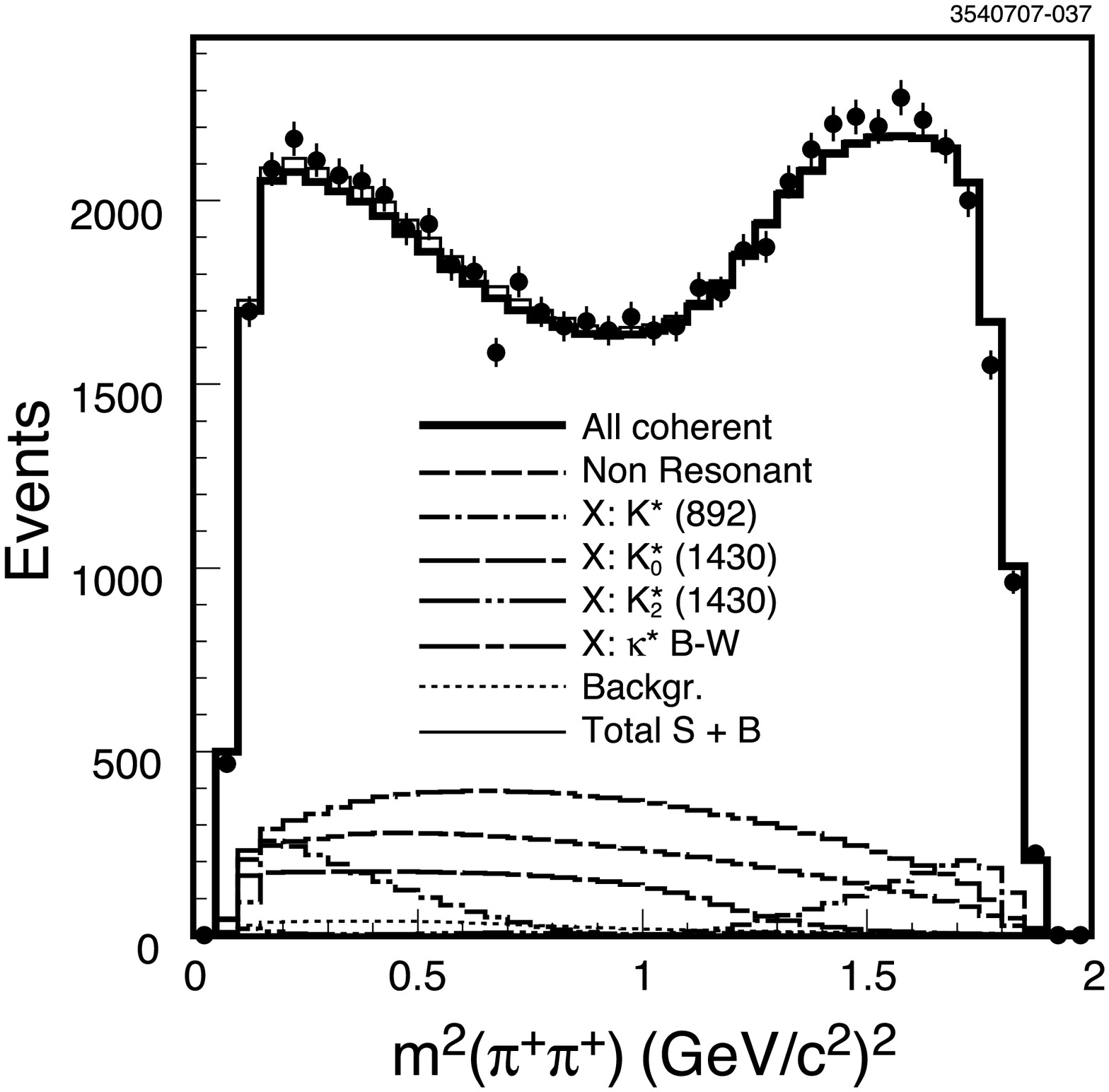}
  \caption{\label{fig:ModelC_Z-proj} The $m^2(\pi^+\pi^+)$ projection in Model C.}
 \end{minipage}
 \begin{minipage}[t]{108mm}
  \includegraphics[width=53mm]{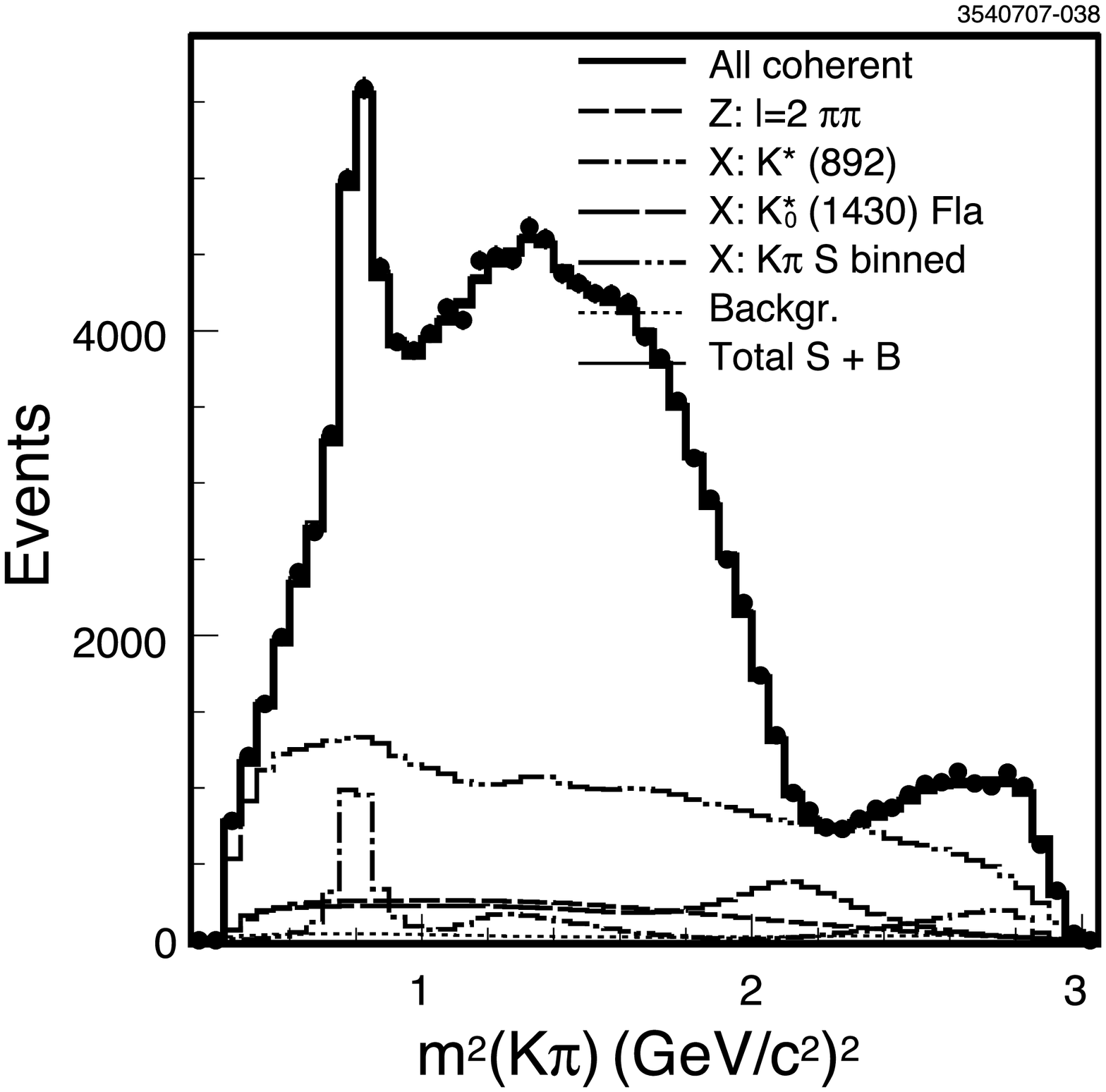}
  \includegraphics[width=53mm]{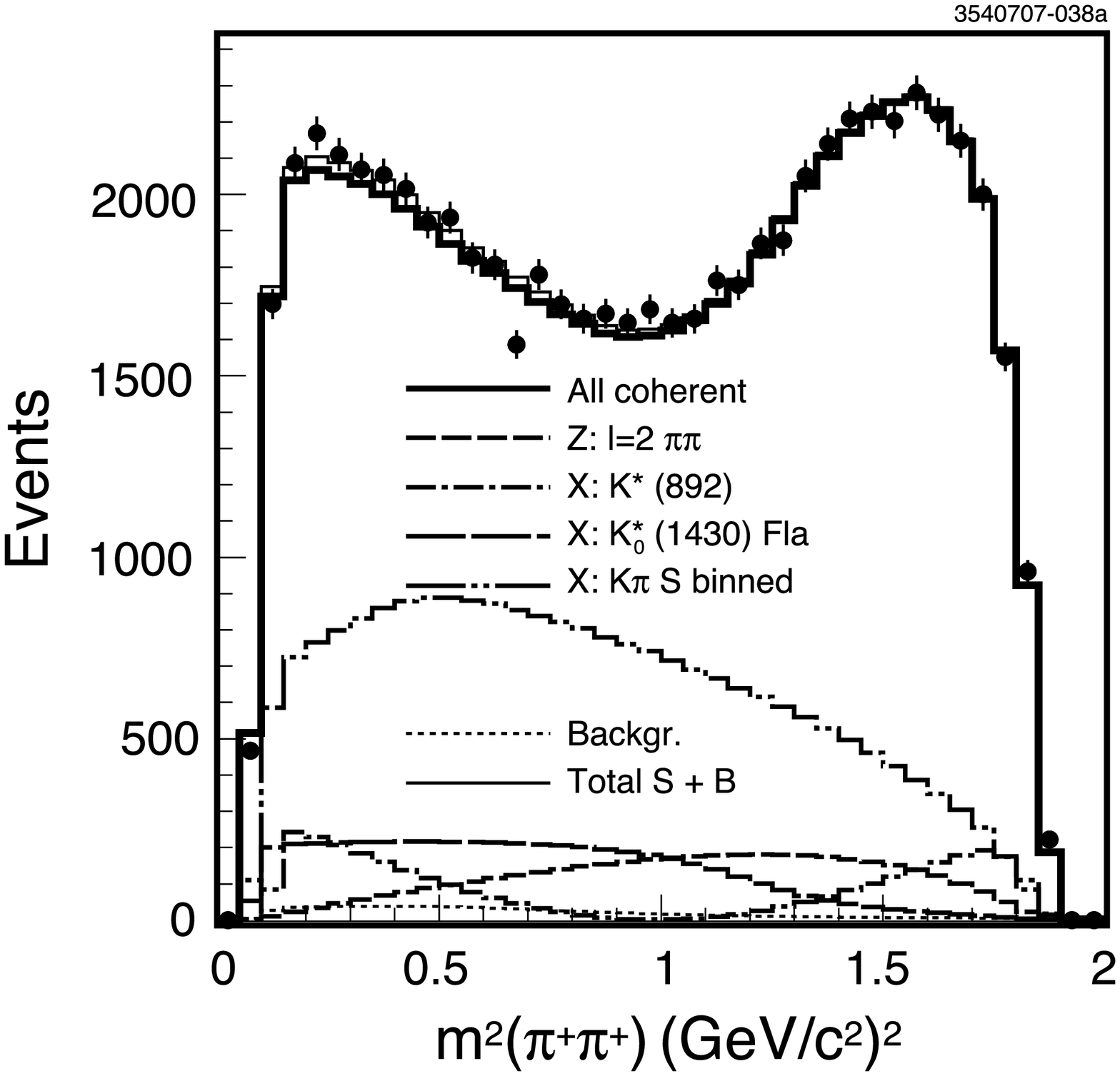}
  \caption{\label{fig:Projections_Model_QMIPWA_I2} The Dalitz plot projections in QMIPWA.}
 \end{minipage}
\end{figure}


\begin{figure}[!htb]
  \includegraphics[width=77mm]{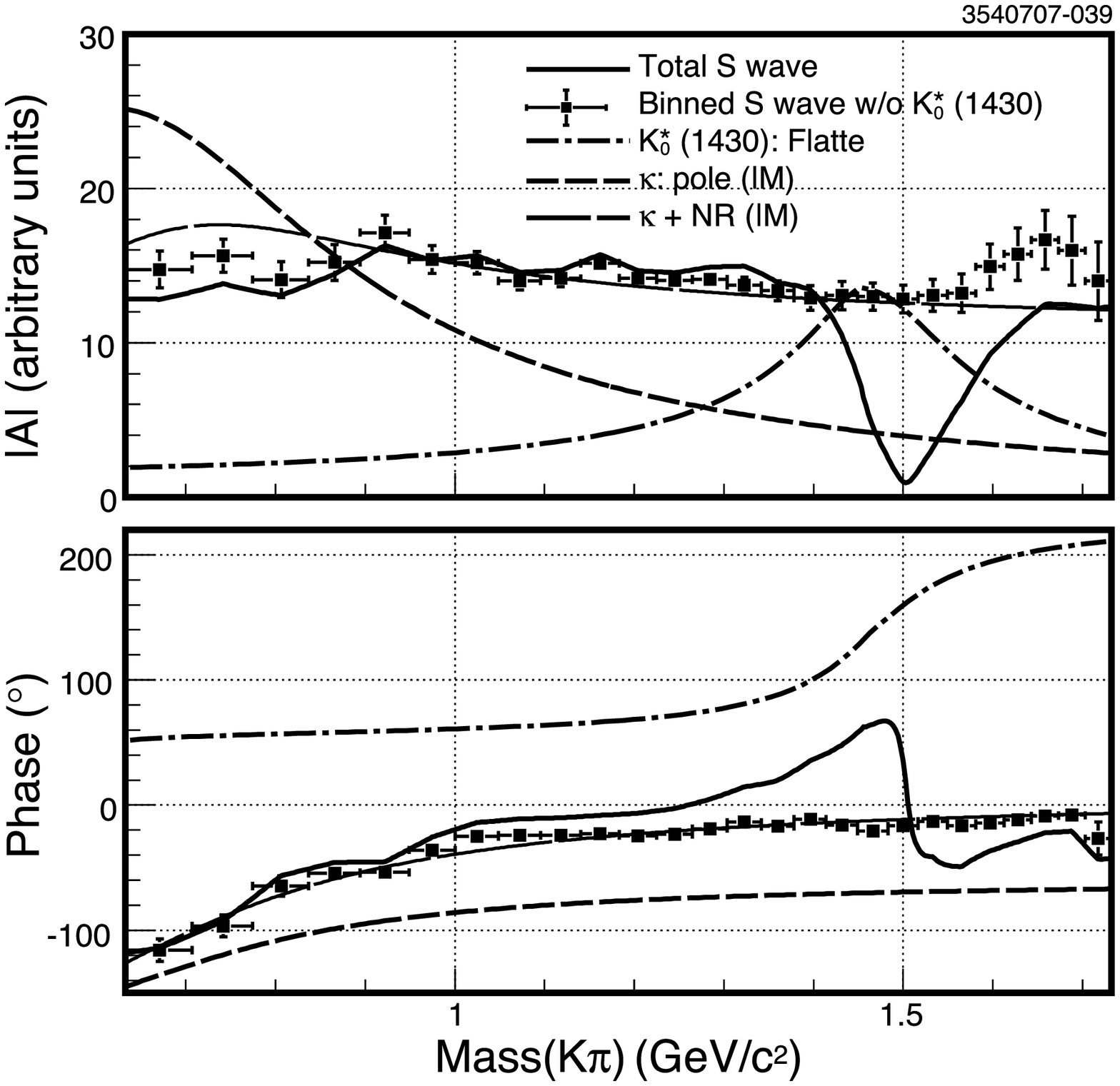}
  \caption{\label{fig:Swave} The $K\pi$ S wave in the isobar model (IM) and QMIPWA.}
\end{figure}

\begin{figure}[!htb]
 \begin{minipage}[t]{77mm}
  \includegraphics[width=77mm]{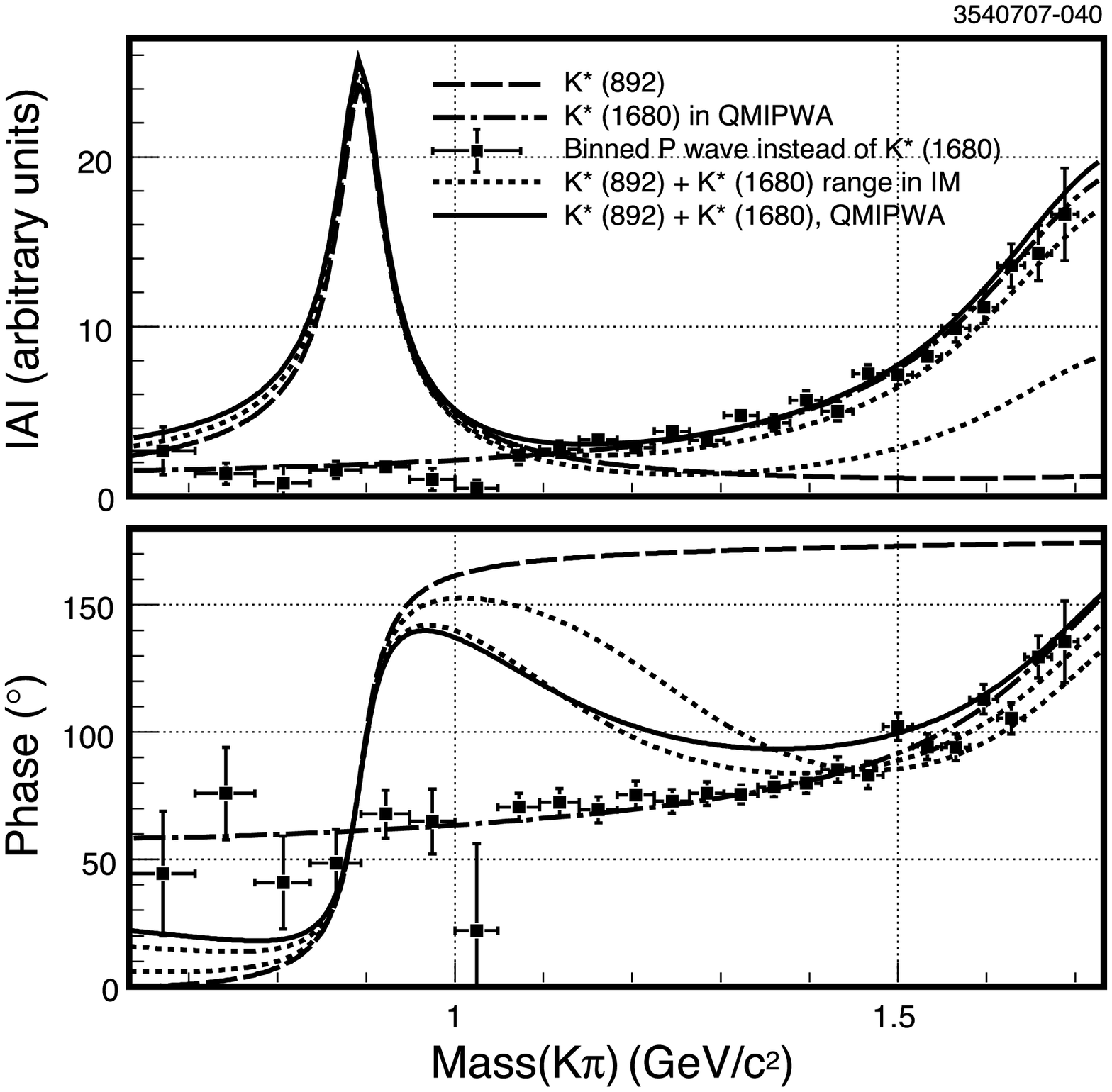}
  \caption{\label{fig:Pwave} The $K\pi$ P wave in the isobar model (IM) and QMIPWA.}
 \end{minipage}
\hfill
 \begin{minipage}[t]{77mm}
  \includegraphics[width=77mm]{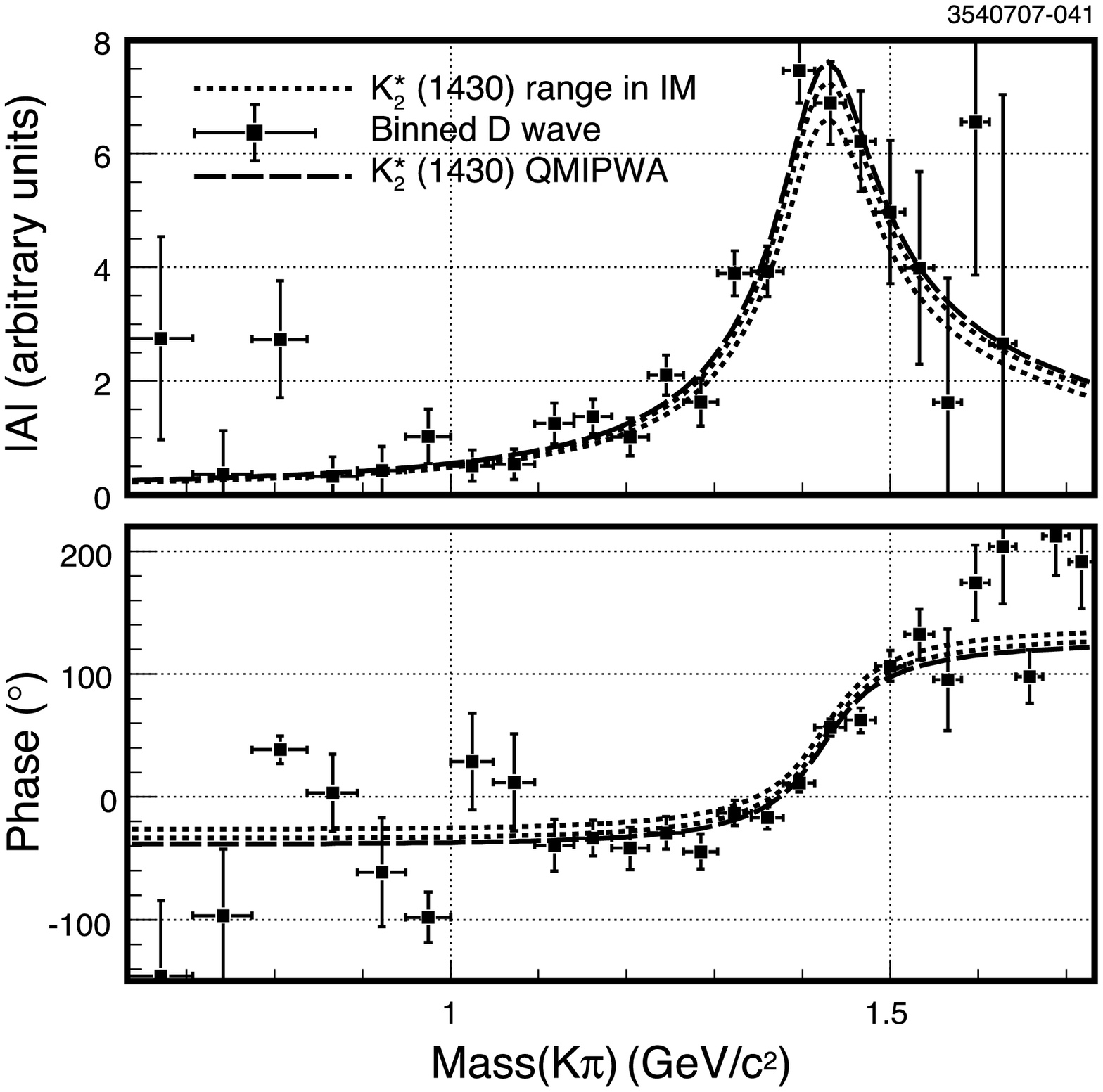}
  \caption{\label{fig:Dwave} The $K\pi$ D wave in the isobar model (IM) and QMIPWA.}
 \end{minipage}
\end{figure}

\begin{figure}[!htb]
 \begin{minipage}[t]{77mm}
  \includegraphics[width=77mm]{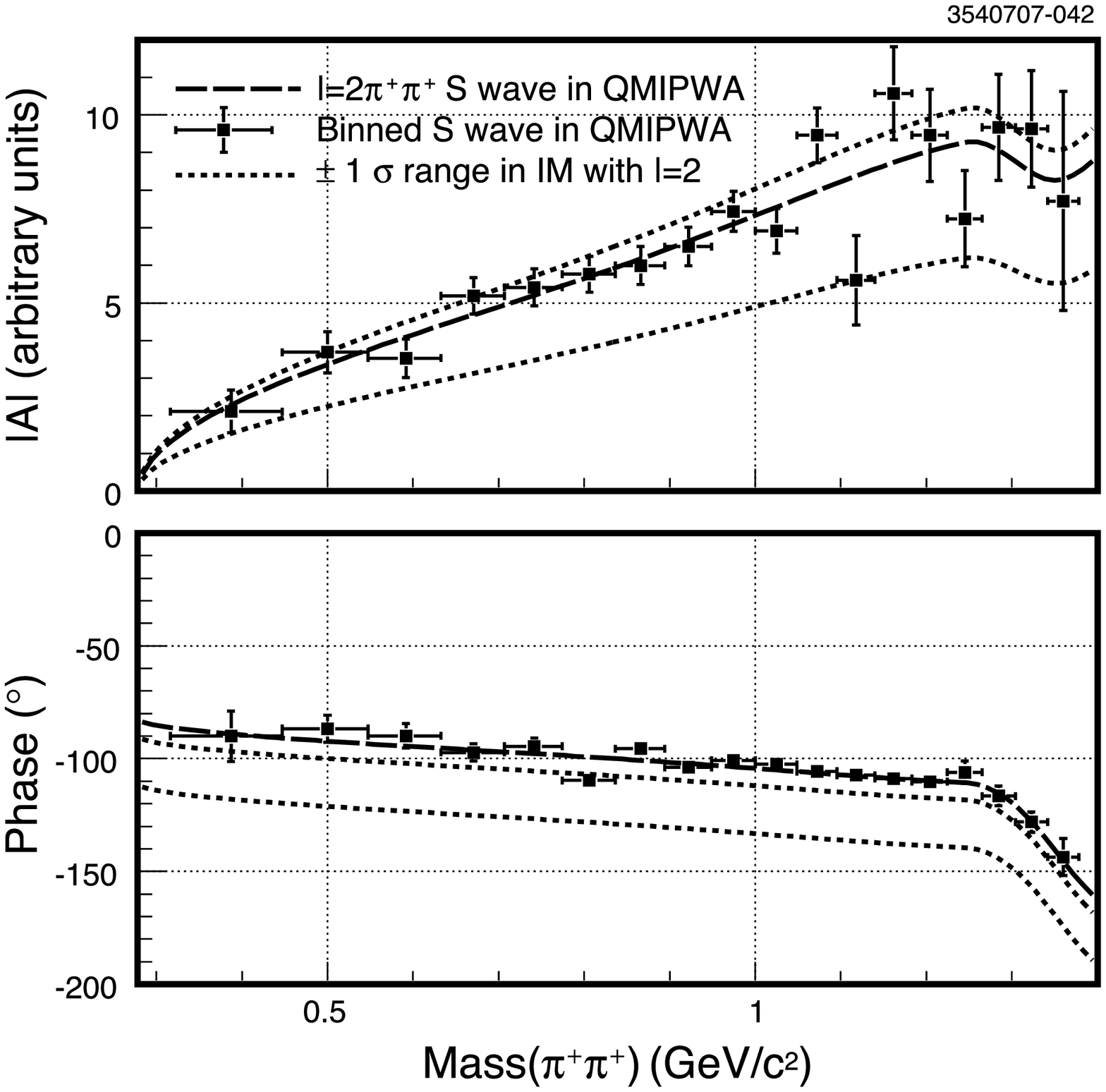}
  \caption{\label{fig:pipiSwave} The I=2 $\pi^+\pi^+$ S wave in the isobar model and QMIPWA.}
 \end{minipage}
\hfill
 \begin{minipage}[t]{77mm}
  \includegraphics[width=77mm]{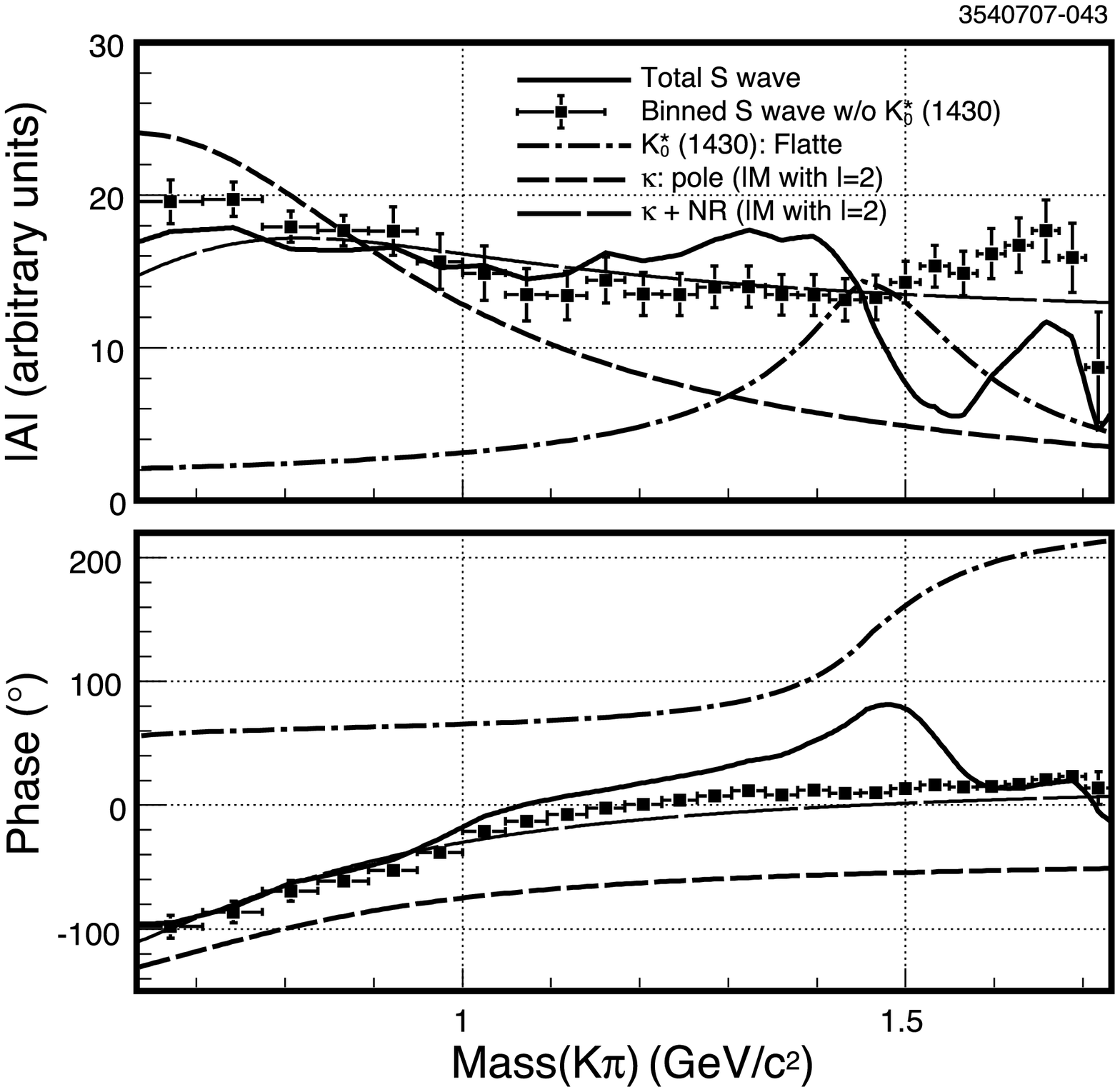}
  \caption{\label{fig:Swave_QMIPWA_I2} The $K\pi$ S wave in the isobar model and QMIPWA 
                  with additional I=2 $\pi^+\pi^+$ S-wave amplitude.}
 \end{minipage}
\end{figure}


In all of our fits that contain only $K\pi$ resonant contributions,
we observe that the p.d.f. significantly deviates the data
in the range $1.3<m^2(\pi^+\pi^+)<1.8$~(GeV/$c^2$)$^2$,
as shown in Fig~\ref{fig:ModelC_Z-proj}.
As a solution to this problem we include an isospin-two
$\pi^+\pi^+$ S-wave contribution in unitary form from 
Ref.~\cite{Achasov_PRD67_2003} which accounts for the
mass-dependent phase 
measured in scattering experiments \cite{Hoodland}.
With its inclusion, we find up to 20\% improvement of the fit 
probability if we allow 
the inelasticity parameter to drop from unity to a much lower value 
above the $\rho-\rho$ threshold starting from $m^2(\pi^+\pi^+)=1.245$~GeV/$c^2$,
using the smooth threshold function, as in Ref.~\cite{Zou_Bugg_2004}.

The framework of the isobar model is good for a first approximation to 
data and comparison with other experiments. 
However, this model violates basic principles, such as unitarity and analyticity.
It is useful to provide a model independent measurement of the partial waves.
E791 in Ref.~\cite{E791_Kpipi} presents the first quasi model-independent partial
wave analysis (QMIPWA) of the $D^+ \to K^-\pi^+\pi^+$ 
decay.  We apply the same technique with minor modifications.

To describe the $K\pi$ amplitude and phase in a model 
independent way we use binned complex amplitudes.
The entire range of $m^2(K\pi)$ from 0.4 to 3~(GeV/$c^2$)$^2$ is 
split into 26 uniform bins, with width 0.1~(GeV/$c^2$)$^2$.
Our fit has 26 amplitudes and 26 phases as fit parameters.
Similarly, the $\pi\pi$ amplitude and phase is defined by 18 bins in 
of $m^2(\pi\pi)$ from 0.1 to 1.9~(GeV/$c^2$)$^2$.
We use linear interpolation between bin centers. 
To alleviate the problem of coarse interpolation we use
Breit-Wigners for parameterization of ``narrow'' resonance structures.
Thus, our S wave is made up of the Flatt\'e function for the
$K_0^*(1430)$ and a binned S-wave amplitude, which replaces the $\kappa$
and non-resonant contributions in the isobar model.
In contrast to E791 we do not use form factors for S-wave contributions.
Our P wave is made up of the Breit-Wigner function for the
$K^*(892)$ and a binned P amplitude, which substitutes for
the wide $K^*(1680)$ Breit-Wigner function in the isobar model.  
Our D wave contains a binned D amplitude which substitutes for
the $K_2^*(1430)$ Breit-Wigner function of the isobar model.
The isospin-two $\pi\pi$ S wave, in addition to the unitary amplitude,
also has a binned amplitude.

The parameters of the binned functions describing the
S, P, and D waves are allowed to float one wave at a time.
Other resonance parameters for the same wave are fixed to their
optimal values from the isobar model. The resonance parameters
of the other spin waves float, as in the isobar model.
First we fit with a floating binned amplitude for the S wave, using the isobar model
approximation for P and D waves. The results of this fit are shown in 
Fig.~\ref{fig:Swave} and Table~\ref{tab:compare_results}.
We find an almost constant S-wave binned amplitude in 
the complete $K\pi$ phase space, which differs slightly
from substituting isobar model components.
The binned phase shows a smooth rise from 
$-120^\circ$ to $-10^\circ$ from minimum to maximum invariant mass.
The total S-wave amplitude is distorted by the 
$K_0^*(1430)$ resonance, as shown by the solid curve.

Next, we fix parameters of the binned S wave and allow the P or 
D waves to float. 
The result of these fits and their comparison with the isobar model are shown in 
Figs.~\ref{fig:Pwave} and \ref{fig:Dwave}, respectively. We find that in both cases 
the P and D binned amplitudes are consistent with the results of the isobar model.
For the D wave amplitude the resolution is poor, 
because its fraction is well below 1\%.

In a fit with floating binned amplitudes for the isospin-two $\pi^+\pi^+$ S wave
we also find a consistent description with the analytical function, shown
in Fig.~\ref{fig:pipiSwave}. 
Adding the isospin-two amplitude in its analytical form the fit with the
$K\pi$ S wave slightly changes the results for the binned $K\pi$ S wave, as shown in
Fig.~\ref{fig:Swave_QMIPWA_I2}.

In systematic cross checks we test the stability of our results by
varying the event selection, efficiency and background parameterization,
variation of model parameters and resonant components, etc.
We expect that systematic uncertainties will be of the same size as the 
statistical errors. The results of this CLEO analysis are preliminary, 
and we plan to publish them after completion of systematic cross checks.

In summary, we present the Dalitz plot analysis of 
the $D^+ \to K^- \pi^+ \pi^+$ decay using currently available CLEO-c data.
Using a simple isobar model we find that our data are consistent with
results obtained in E791 experiment~\cite{E791-2002}.
Inclusion of the isospin-two $\pi^+ \pi^+$ S wave improves 
consistency of this model with data. Finally, we apply quasi model-independent
partial wave analysis \cite{E791_Kpipi} and measure 
the amplitude and phase of the $K\pi$ and  $\pi^+\pi^+$ S waves in the range
of invariant masses from the threshold to the maximum in this decay.

We gratefully acknowledge the effort of the CESR staff
in providing us with excellent luminosity and running conditions.
This work was supported by
the A.P.~Sloan Foundation,
the National Science Foundation,
the U.S. Department of Energy, and
the Natural Sciences and Engineering Research Council of Canada.


\end{document}